\documentclass[fleqn,usenatbib]{mnras}

\usepackage{newtxtext,newtxmath}

\usepackage[T1]{fontenc}

\DeclareRobustCommand{\VAN}[3]{#2}
\let\VANthebibliography\thebibliography
\def\thebibliography{\DeclareRobustCommand{\VAN}[3]{##3}\VANthebibliography}

\usepackage{graphicx}	
\usepackage{amsmath}	

\title[Free Energy of Anisotropic Strangeon Stars]{Free Energy of Anisotropic Strangeon Stars}

\author[Chen et al.]{
Shichuan Chen,$^{1,2}$\thanks{E-mail: 2001110255@stu.pku.edu.cn}
Yong Gao,$^{1,2}$\thanks{E-mail: gaoyong.physics@pku.edu.cn}
Enping Zhou$^{3}$\thanks{E-mail: ezhou@hust.edu.cn}
and Renxin Xu$^{1,2}$\thanks{E-mail: r.x.xu@pku.edu.cn}
\\
$^{1}$Department of Astronomy, School of Physics, Peking University, Beijing 100871, China\\
$^{2}$Kavli Institute for Astronomy and Astrophysics, Peking University, Beijing 100871, China\\
$^{3}$Department of Astronomy, School of Physics, Huazhong University of Science and Technology, 1037 Luoyu Road, Wuhan, 430074, China
}

\date{Accepted XXX. Received YYY; in original form ZZZ}

\pubyear{2024}

\begin{document}
\label{firstpage}
\pagerange{\pageref{firstpage}--\pageref{lastpage}}
\maketitle

\begin{abstract}
Can pulsar-like compact objects release further huge free energy besides the kinematic energy of rotation? This is actually relevant to the equation of state of cold supra-nuclear matter, which is still under hot debate. Enormous energy is surely needed to understand various observations, such as $\gamma-$ray bursts, fast radio bursts and soft $\gamma-$ray repeaters. In this paper, the elastic/gravitational free energy of solid strangeon star is revisited for strangeon stars, with two anisotropic models to calculate in general relativity. It is found that huge free energy (> $10^{46}$ erg) could be released via starquakes, given an extremely small anisotropy ($(p_{\rm t}-p_{\rm r})/p_{\rm r} \sim 10^{-4}$, with $p_{\rm t}$/$p_{\rm r}$ the tangential/radial pressure), implying pulsar-like stars could have great potential of free energy release without extremely strong magnetic fields in solid strangeon star model.
\end{abstract}




\section{Introduction}           
\label{sect:intro}

A compact object composed of dense matter at supra-nuclear density forms after stopping the release of nuclear free energy in massive stars, which was initially termed as ``gigantic nucleus'' by~\cite{1932PhyZS...1..285L}.
Can this kind of compact stars release further huge free energy besides the rotational energy? This is an issue with a long history, relevant to the state equation of cold supra-nuclear matter, which is still challenging in both physics and astronomy nowadays~\citep{2023AN....34430008X}.

Observationally, an evolution of a post-burst relativistic fireball with free energy injection from the compact star through magnetic dipole radiation may provide a natural explanation for  the plateau of $\gamma$-ray bursts (GRBs)~\citep{1998A&A...333L..87D,2001ApJ...552L..35Z,2022Natur.612..236M}.
As the companion piece of GRBs, fast radio bursts (FRBs), especially the repeating ones with  high burst rate, are calling for enormous free energy of compact central engines, which are most likely pulsar-like objects~\citep{2018ApJ...852..140W,2022SCPMA..6589511W,2020Natur.586..693L,2021Natur.598..267L,2022Natur.609..685X}.
In addition, tremendous free energy is shown in the observations of the flares of galactic even extra-galactic sources, so-called soft $\gamma$-ray repeaters, especially for the giant ones~\citep{2005Natur.434.1098H,2020Natur.587...54C,2021NatAs...5..385F}, with extremely bright giant flares with energy of $10^{44-47}$ erg \citep{1999Natur.397...41H, 2005Natur.434.1107P}.

Theoretically, though the possibility of a solid core~\citep{1972ARA&A..10..427R,1973NPhS..243...63C} cannot yet be ruled out, a conventional neutron star (NS) is fluid-like except for a solid crust (i.e., similar to a raw egg), the free energy of which could be negligible, but it might be significant in case of a state strongly magnetized~\citep{1992ApJ...392L...9D,1992Natur.357..472U,1993ApJ...408..194T}, so-called magnetars~\citep{1995MNRAS.275..255T, 1998Natur.393..235K} with extremely strong magnetic fields ($\sim10^{13-15}$G). It seems that the theory of magnetars has been successful to explain many observations of anomalous X-ray pulsars and soft $\gamma-$ray repeaters, e.g., the energy budgets and the braking indices \citep{2021AN....342..369G,2020Univ....6...63W,2016MNRAS.456...55G,2017ApJ...849...19G}.
Nevertheless, nucleon-like units with strangeness, called strangeons, may form in bulk supra-nuclear matter produced during core-collapse supernova, and a strangeon star (SS)~\citep{2003ApJ...596L..59X,2009MNRAS.398L..31L,2023AdPhX...837433L} should be in a globally solid state (i.e., similar to a cooked egg) due to the large masses of and the strong coupling between strangeons.
A calculation of the free energy for anisotropic SS was presented in Newtonian gravity, showing a huge
amount of energy released via starquakes when stellar stresses reach a critical value~\citep{2006MNRAS.373L..85X}, and an updated version with Einstein's gravity will be given in the present work.

Although it's a common assumption in studying pulsar-like compact objects that the pressure is isotropic, it may not be true since some processes may induce anisotropy because of, for instance, strong magnetic field \citep[e.g.][]{2001ApJ...554..322C, 2012MNRAS.427.3406F, 2013MNRAS.435L..43C, 2015AN....336..866G}, relativistic nuclear interaction \citep{1972ARA&A..10..427R, 1974ARA&A..12..167C}, pion condensation \citep{1972PhRvL..29..382S}, phase transitions \citep{1998NuPhB.531..478C}, superfluid core \citep{2000PhR...328..237H}, and so on. However, it's quite difficult to compute the exact anisotropic models on physical ground from first principle. Several heuristic anisotropic models have been put forward \citep[e.g.][]{1974ApJ...188..657B, 2013PhRvD..88h4022H}, based on some assumptions to make the models physically acceptable and available.

Additionally, the free energy of pulsar-like compact object depends certainly on the equation of state of bulk matter at supra-nuclear density, and it is generally thought that strangeness would play an important role in understanding the puzzling state, to be probably the first big problem solved in the era of gravitational-wave astronomy~\citep{1971PhRvD...4.1601B,1984PhRvD..30..272W}, see also~\cite{2018SCPMA..61j9531X} for a brief introduction.
It is then suggested that pulsars could be strange quark stars (QSs), having similar mass and radius to that of normal NSs ~\citep{1986A&A...160..121H,1986ApJ...310..261A}, which makes QS a possible candidate model for this kind of compact objects. It is worth noting that the basic units of a strange star would be quarks for a QS, but could be strangeons if three-flavored quarks are localized in strangeons as for nucleons in the two-flavored case \citep{2003ApJ...596L..59X}. The model of SS has been successful to explain many phenomena of pulsar-like stars, including the subpulse-drifting \citep{1999ApJ...522L.109X,2019SCPMA..6259505L}, the glitches interpreted with star-quakes \citep{2004APh....22...73Z,2014MNRAS.443.2705Z,2018MNRAS.476.3303L,2021MNRAS.500.5336W,2023MNRAS.520.4289L}, the Optical/UV excess in X-ray dim isolated NSs \citep{2017ApJ...837...81W}, as well as massive pulsars ($\sim\ \rm2M_\odot$) proposed before discoveries~\citep{2009MNRAS.398L..31L}.
Recently, the SS model is also consistent with the results of tidal deformability~\citep{2019EPJA...55...60L} of and the light curve \citep{2018RAA....18...24L} from GW170817. In addition, photon-driven mechanism might alleviate the current difficulty in core-collapse supernovae by forming a strange star inside the collapsing core~\citep{2007ApJ...668L..55C}, producing more free energy injected into explosive shock wave than that of conventional neutrino-driven ones~\citep{2015ApJ...801L..24M}. The model could also be tested in the future by detecting gravitational-wave echos associated with SSs \citep{2023PhRvD.108f3002Z}. In summary, there are many differences between the SSs and normal NSs, not only in surface features, but also the global properties such as maximal mass and tidal deformability. It's expected to see these differences in the future observations (see the review \cite{2017JPhCS.861a2027X,2023AdPhX...837433L} and references therein).

The free energy of solid SSs is focused in the paper, with numerical calculations of the strain energy release during a starquake within general relativity in a spherically symmetric spacetime. This paper is motivated by \cite{2006MNRAS.373L..85X}, which showed in Newtonian gravity that a solid pulsar can release a large amount of free energy from elastic or gravitational energy during a starquake due to the anisotropy of the solid star. We calculate here this kind of free energy with more physically acceptable anisotropic models and EOS in Einstein's gravity. It is evident from our calculation that the huge free energy release of anisotropic solid SSs can naturally provide an alternative way to power $\gamma$-ray bursts, fast radio bursts and soft $\gamma$-ray repeaters without extremely strong magnetic fields. Such kind of stress-energy stored in anisotropic stars to be releasable during a starquake has also been emphasized by \cite{2023InJPh..97.3379K}, showing how the difference between sound propagation in radial and tangential directions would be used to identify potentially stable regions within a configuration.

The solid type stars have advantage over fluid type ones (e.g. conventional NS) in releasing the free energy from starquakes, since in the fluid-like star case, the starquake can only happen in the outer crust, while for the solid star, the whole star can release free energy by starquakes. This is one of the reason we choose the model of SS, one type of solid strange quark stars, rather than other fluid-like stars.
Since we mainly focus on the difference of free energy of different parameters, we ignore the influence of anisotropy on the shape, structure, and radiation of the star. We think it's reasonable because the case we study has only small anisotropy ($(p_{\rm t}-p_{\rm r})/p_{\rm r} \leq 10^{-4}$, with $p_{\rm t}$/$p_{\rm r}$ the tangential/radial pressure), which should have minor influence on the results. We also neglect the influence of rotation of the star, since it has tiny influence on the gravitational mass (hence the free energy) of SS in the case \citep{2022MNRAS.509.2758G}.

The paper is arranged as followed. In section 2, we will introduce the methods and models used to calculate the free energy of SSs in the anisotropic case, including the modified TOV equations in subsection 2.1, two anisotropic models in subsection 2.2, the equation of state of SS in subsection 2.3, the method to calculate the free energy in subsection 2.4 and the main results of our calculations in subsection 2.5. We make conclusions and discussions in Section 3. We will use cgs system of units throughout the paper.

\section{Methods and Models}
\label{sect:model}

\subsection{TOV equations in the anisotropic case}
\label{sec:aniso}
For a spherically symmetric star modelled by perfect fluid in static equilibrium, the Tolman-Oppenheimer-Volkoff (TOV) equations constrain the structure of the star. But the isotropic star is only a common assumption. It's natural to believe that strongly interacting matter such as NSs should be described by locally anisotropic equation of state (EOS) \citep[e.g.][]{1972ARA&A..10..427R,1974ApJ...188..657B}.

For simplicity, consider a static distribution of anisotropic matter in spherically symmetric spacetime. In Schwarzschild-like coordinates, the metric can be written as:
\begin{equation}
    ds^2 = -e^{2\alpha(r)}c^2dt^2 + e^{2\beta(r)}dr^2 + r^2(d\theta^2+\sin^2\theta d\phi^2)
	\label{eq:schmetic}
\end{equation}
The spherical symmetry spacetime also implies that the stress-energy tensor $T_{\mu\nu}$ can be written as \citep{3-20220092}
\begin{equation}
    T_{\rm\mu\nu}=(\rho + p_{\rm t}/c^2) u_{\mu}u_{\nu} + p_{\rm t}g_{\mu\nu} + ( p_{\rm r} - p_{\rm t} ) \theta_{\mu}\theta_{\nu}
	\label{eq:tmunu}
\end{equation}
where $\rho$ is the energy density, $p_{\rm r}$ is the radial pressure , $p_{\rm t}$ is the tangential pressure, $u^{\mu}$ is the unit 4-velocity of the matter, $u_{\mu}=g_{\mu\nu}u^{\nu}$, and $\theta^{\mu}$ is the unit space-like vector in the direction of radial vector, $\theta_{\mu}=g_{\mu\nu}\theta^{\nu}$.

Combining with the Einstein equations, we have \citep{1974ApJ...188..657B}
\begin{equation}
    e^{2\beta(r)}=(1-\frac{2Gm(r)}{rc^2})^{-1}
	\label{eq:tov1}
\end{equation}
\begin{equation}
    \frac{d\alpha}{dr}=e^{2\beta(r)}\frac{G}{c^4r^2}(m(r)c^2+4\pi r^3p_{\rm r})
	\label{eq:tov2}
\end{equation}
\begin{equation}
	\frac{dp_{\rm r}}{dr}=-(p_{\rm r}+\rho c^2)\frac{d\alpha}{dr}+\frac{2\Pi}{r}
	\label{eq:tov3}
\end{equation}
where $\Pi=p_{\rm t}-p_{\rm r}$ measures the local anisotropy and $m(r)=\int^r_04\pi r^2 \rho dr$ is the mass within the radius $r$.

Equation \eqref{eq:tov1}, \eqref{eq:tov2}, \eqref{eq:tov3} are the generalized TOV equations in the anisotropic case. Compared to the normal TOV equations,  equation \eqref{eq:tov3} shows that the difference comes from the new variable $\Pi=p_{\rm t}-p_{\rm r}$, which should be determined by a new relation assumed, explained in \S~2.2. In this paper, we apply these equations to solid SSs. We think it's reasonable since the SSs with small anisotropy can be approximated by anisotropic fluid.

\subsection{Two anisotropic models}

One of the most important issues is how to determine the model of $\Pi$, the difference between $p_{\rm t}$ and $p_{\rm r}$. Since it's very difficult to obtain $\Pi$ on physical grounds from first principle, one could only guess some heuristic models. We assume the anisotropy is small so it doesn't change structures a lot, and the EOS only depends on $p_{\rm r}$ not $p_{\rm t}$.

There are some minimal conditions on $\Pi$ to make solutions physically
acceptable \citep{2018EPJC...78..673E}. In a nutshell, these conditions include: the interior solution should match continuously to the exterior Schwarzschild solution; the metric functions must be finite and non-zero within the star; the density and pressure must be non-negative and finite everywhere, and must be monotonic decreasing with radius; the radial and tangential pressure at the origin must be the same; the energy conditions should be satisfied; the causality condition must be satisfied within the star, i.e. the speed of sound must be lower than the light speed.

For simplicity, we choose two models of $\Pi$ in our calculation. The first model is $\Pi=-\eta_1 R_1 \frac{dp_{\rm r}}{dr}$, where $\eta_1$ is a dimensionless constant, and $R_{1}$ is a constant with the dimension of length, to be $R_{1} = 10\ {\rm km}$ for the typical radius of pulsars. The second one is the HB model with $\Pi=-\eta_2 r \frac{dp_{\rm r}}{dr}$~\citep{2013PhRvD..88h4022H}, where $\eta_2$ is a dimensionless constant too. The constants $\eta_1$ and $\eta_2$ measure the anisotropy of the star, $\eta_{1,2}=0$ implies that the star is isotropic and has no strain energy. Both the models satisfy the conditions above and are physically acceptable.

\subsection{Equation of state of strangeon matter}

We choose the phenomenological Lennard-Jones model of SSs \citep{2009MNRAS.398L..31L}, which assumes an interaction potential between two strangeons of
\begin{equation}
	u(r)=4\epsilon[(\frac{\sigma}{r})^{12}-(\frac{\sigma}{r})^{6}]
	\label{eq:lj}
\end{equation}
where $\epsilon$ is a constant that represents the depth of the potential, $\sigma$ is a constant that represent the distance between two strangeons when their interaction potential $u(r)$ is zero.

The Lennard-Jones model was usually used as the interaction between molecules, with the property of long-range attraction and short-range repulsion. The lattice QCD show that there is a strong repulsive core of a few hundred MeVs at short distances ($r\le0.5\ {\rm fm}$) surrounded by an attractive well at medium and long distances \citep{2007PhRvL..99b2001I,2007Natur.445..156W}. This kind of potential helps quark matter crystallize and form solid strange stars.

If we adopt the simple cubic lattice structure, ignore the surface tension and vibration energy (since it's small compared to the potential energy and the rest energy), the total energy density $\epsilon_q$ and pressure $p$ of strangeon matter can be calculated as \citep{2009MNRAS.398L..31L}
\begin{equation}
	\epsilon_q=\epsilon_p+nN_qm_qc^2=2\epsilon(A_{12}\sigma^{12}n^5-A_6\sigma^6n^3)+nN_qm_qc^2
	\label{eq:energydensity}
\end{equation}
\begin{equation}
	p=n^2\frac{d(\epsilon_q/n)}{dn}=4\epsilon(2A_{12}\sigma^{12}n^5-A_6\sigma^6n^3)
	\label{eq:pressure}
\end{equation}
where $A_{12} = 6.2$, $A_6 = 8.4$, which are constants got from the simple cubic structure, $N_q$ is the number of quarks in one strangeon, $m_q$ is the quark mass which assumed to be one-third of the nuclear mass and $n$ is the number density of strangeons.

\begin{figure}
	\includegraphics[width=0.8\columnwidth]{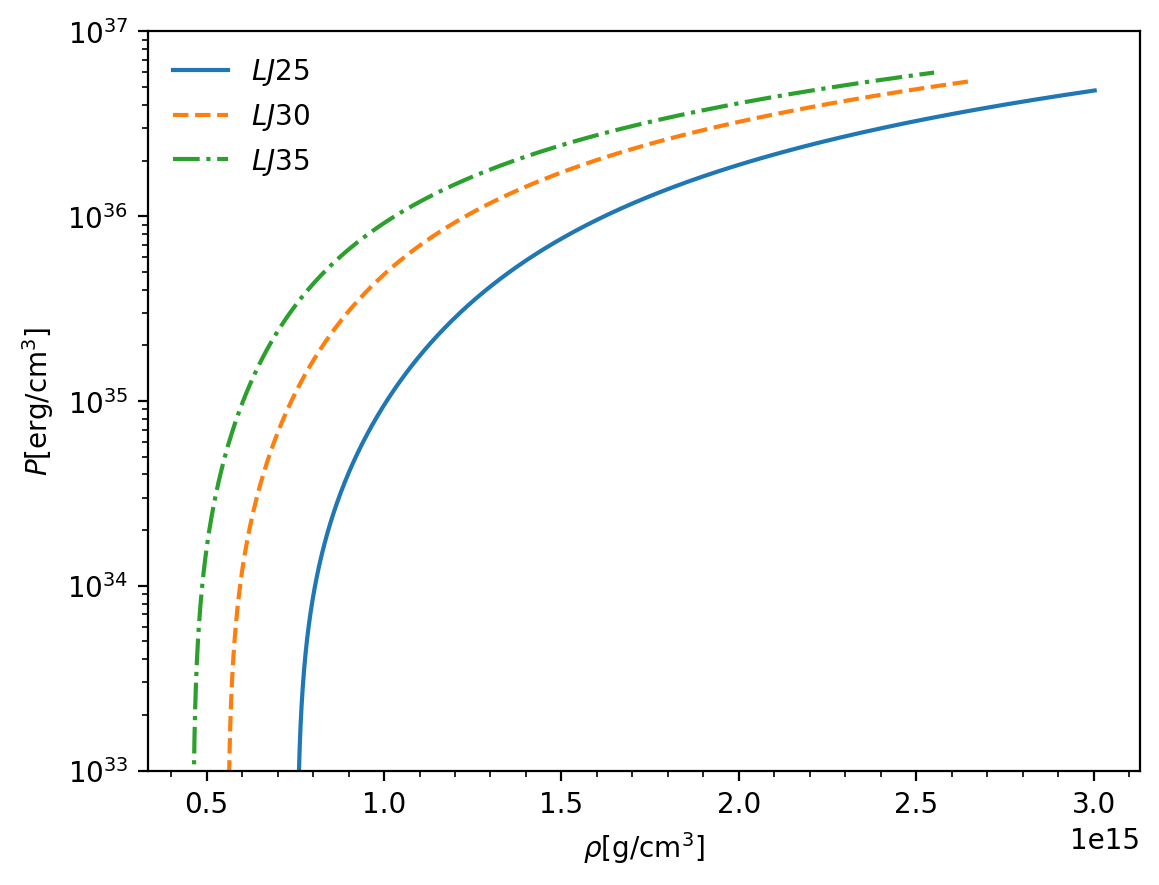}
    \caption{The P-$\rho$ diagram of SS for different Lennard-Jones SS model adopted in the paper.}
    \label{fig:P-rho_LJ}
\end{figure}

\begin{figure}
	\includegraphics[width=0.8\columnwidth]{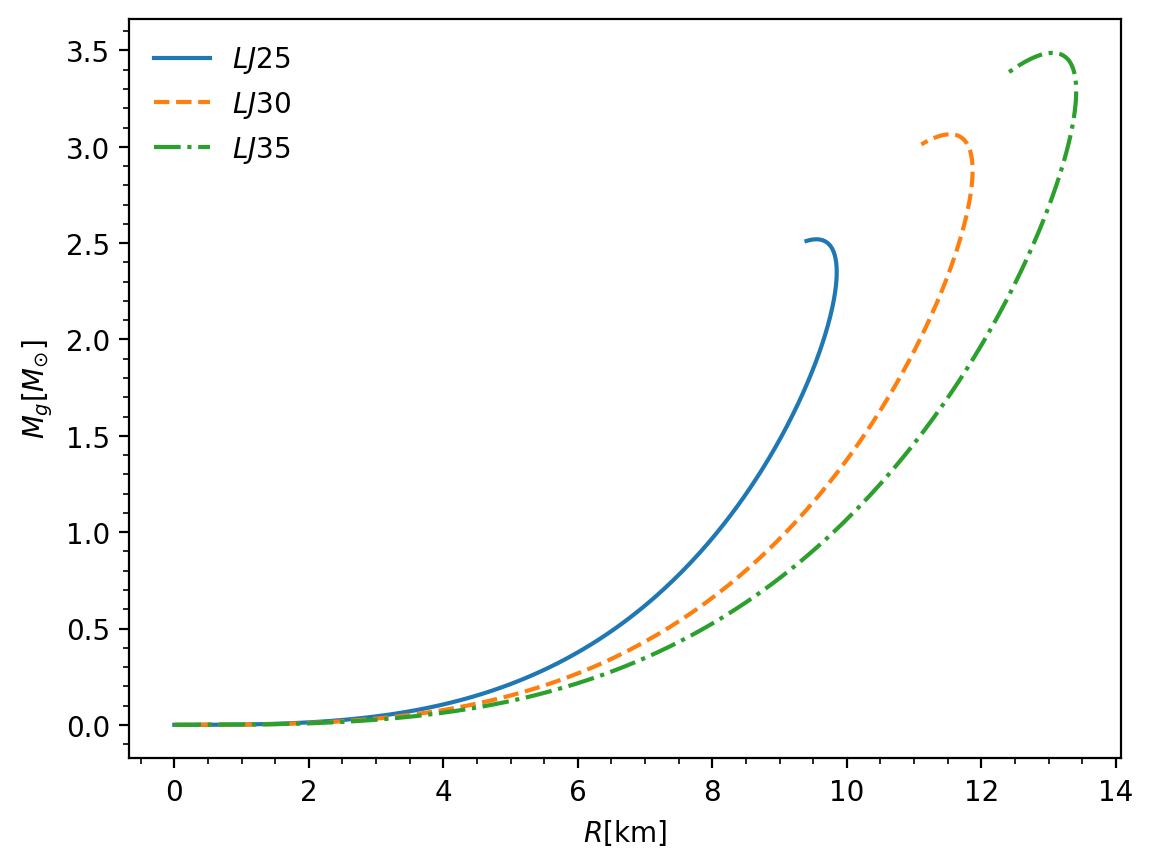}
    \caption{The M-R diagram of SS for different Lennard-Jones SS model adopted in the paper.}
    \label{fig:M-R_LJ}
\end{figure}

\begin{figure}
	\includegraphics[width=0.8\columnwidth]{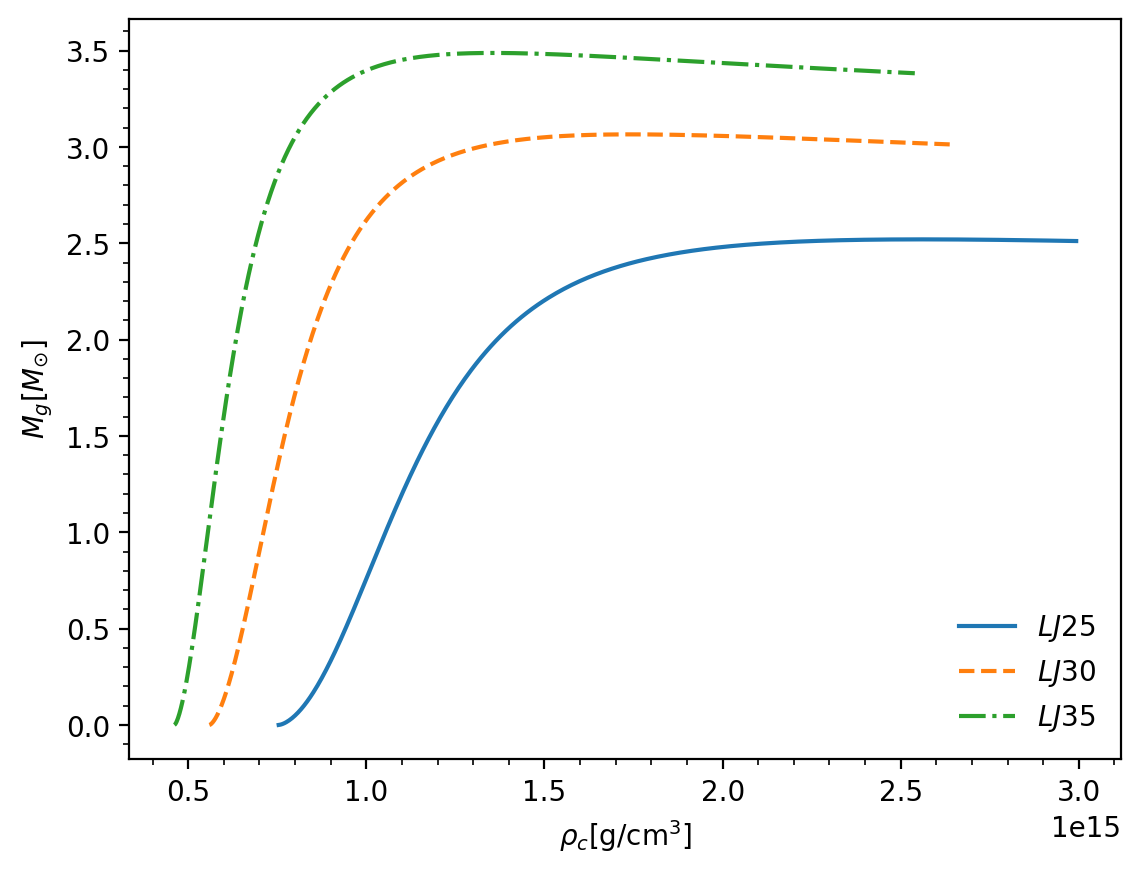}
    \caption{The M-$\rho_c$ diagram of SS for different Lennard-Jones SS model adopted in the paper.}
    \label{fig:M-rho_LJ}
\end{figure}

We adopt three groups of parameters of this Lennard-Jones SS model, which named after their maximum gravitational mass. We choose the maximum gravitational mass $M_{max}$ to be around $2.5,\ 3.0,\ 3.5\ M_{\odot}$ by setting different values of $\sigma$ and $\epsilon$. These parameters are listed in table~\ref{tab:EOS}, where $n_s=(A_6/2A_{12})^{1/2}N_q/3\sigma^3$ is the surface number density of baryons, $N_q$, $\sigma$ and $\epsilon$ are the same parameters as that in equation \eqref{eq:lj} and \eqref{eq:energydensity}. $N_q$ is set to be 18, because a strangeon of 18-quark cluster has maximum symmetry, being completely asymmetric in spin, flavor and color space \citep{1988PhRvL..60..677M}. The relation between the pressure $P$ and energy density $\rho$ for these three models is shown in Figure \ref{fig:P-rho_LJ}. The relation between the gravitational mass $M_g$ and the radius $R$ is shown in Figure \ref{fig:M-R_LJ}, and the relation between $M_g$ and the central energy density $\rho_c$ is shown in Figure \ref{fig:M-rho_LJ}. The radius of SS with maximum mass in model LJ25, LJ30, LJ35 is 9.55 km, 11.54 km and 13.06 km respectively. The adiabatic sound speed of SSs has been discussed in \cite{2009MNRAS.398L..31L}. This model of SS is potentially stable against cracking if $-1 \le v_{\rm r}^2-v_{\rm t}^2 = \frac{\partial p_{\rm r}}{\partial \rho} - \frac{\partial p_{\rm t}}{\partial \rho} = -\frac{\partial \Pi}{\partial \rho} \le 0$ \citep{2007CQGra..24.4631A,2015JPhCS.600a2014G}, which is satisfied almost everywhere within the star.

Though it has been reported that a black hole (BH) with mass less then 3.5 $M_{\odot}$ was found \citep{2019Sci...366..637T}, we think the model LJ35 is still meaningful. Since the 2 $\sigma$ confidence interval of the BH mass found in \cite{2019Sci...366..637T} is from 2.6 to 6.1 $M_{\odot}$, it's still uncertain whether model LJ35 has exceeded minimum mass of BH or not. And from the point of mathematics, it's acceptable even if 3.5 $M_{\odot}$ exceeds the upper limit of NS mass a bit, since we mainly focus on the influence of different parameters on the free energy.

\begin{table}
	\centering
	\caption{The parameters of Lennard-Jones SS model we used in the calculation.}
	\label{tab:EOS}
	\begin{tabular}{lcccr} 
		\hline
		name & $n_{\rm s}\ [{\rm fm}^{-3}]$ & $N_{\rm q}$ & $\epsilon$ [MeV] & $M_{\rm max}$\\
		\hline
		LJ25 & 0.48  & 18    & 20  & $2.5M_{\odot}$\\
		LJ30 & 0.36  & 18    & 30  & $3.0M_{\odot}$\\
		LJ35 & 0.30  & 18    & 40  & $3.5M_{\odot}$\\
		\hline
	\end{tabular}
\end{table}

\subsection{To calculate the free energy}

With the generalized TOV equations \eqref{eq:tov1}, \eqref{eq:tov2}, and \eqref{eq:tov3} in the anisotropic case, Lennard-Jones SS EOS \eqref{eq:energydensity}, \eqref{eq:pressure} and the choice of anisotropic model, either $\Pi=-\eta_1 R_1 \frac{dp_{\rm r}}{dr}$ or $\Pi=-\eta_2 r \frac{dp_{\rm r}}{dr}$, we have the  complete equations to solve out the whole system. Once given the central energy density $\rho_c$, one can integrate the generalized TOV equations from the center to the surface, and obtain the radius, the gravitational mass $M_{\rm g}$ and the baryon mass $M_{\rm b}$ of the SS, which can be calculated as
\begin{equation}
	M_{\rm g}=\int^R_04\pi \rho r^2 dr
	\label{eq:massg}
\end{equation}
\begin{equation}
	M_{\rm b}={\rm 930 MeV}/c^2 \int^R_04\pi n r^2 e^{\beta(r)} dr
	\label{eq:massb}
\end{equation}

As SS spins down, the centrifugal force will decreases, and elastic energy will accumulate to resist the deformation of the star. When the elastic energy exceeds a certain value, the star can no longer stand against it, and a starquake occurs. This kind of earthquakes doesn't change star's volume. However, a solid star may have a starquake in case of accretion, which can change it's volume. The two types of starquake models of SSs and their relation to glitches and pulsar's spin down have been discussed in \cite{2014MNRAS.443.2705Z}.

The binding energy of the star can be calculated as $E_{\rm b}= (M_{\rm b} - M_{\rm g})c^2$. Starquakes may cause the sudden change of $\Pi$, with a release of the gravitational energy as well as the strain energy. The difference of binding energy ${\rm\Delta} E_{\rm b}=E_{\rm b}(\eta_{1,2})-E_{\rm b}(\eta_{1,2}=0)$ between the star with $\eta_{1,2}\neq0$ and $\eta_{1,2}=0$ may imply the free energy the star can release during the starquakes.

\subsection{Results}

\begin{figure}
	\includegraphics[width=0.9\columnwidth]{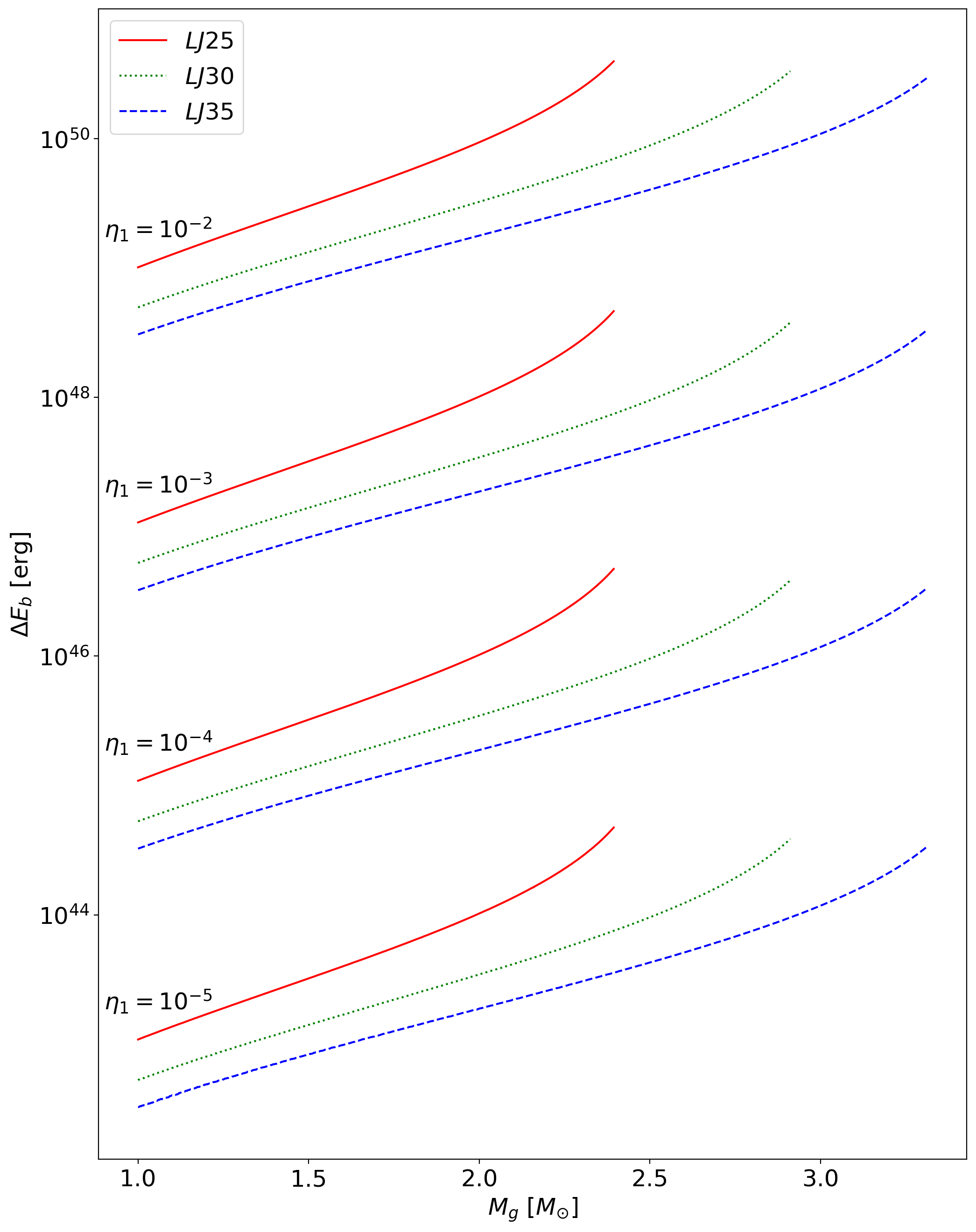}
    \caption{The difference of binding energy as a function of gravitational mass with the anisotropic model $\Pi=-\eta_1R_1dp_{\rm r}/dr$. Three different line styles (or colors) correspond to three choices of EOS, which are listed in table~\ref{tab:EOS}. The lines in the same line style (or colors) from top to bottom have different $\eta_1$, which are labelled in the graph.}
    \label{fig:n=0_deltaEb}
\end{figure}

\begin{figure}
	\includegraphics[width=0.9\columnwidth]{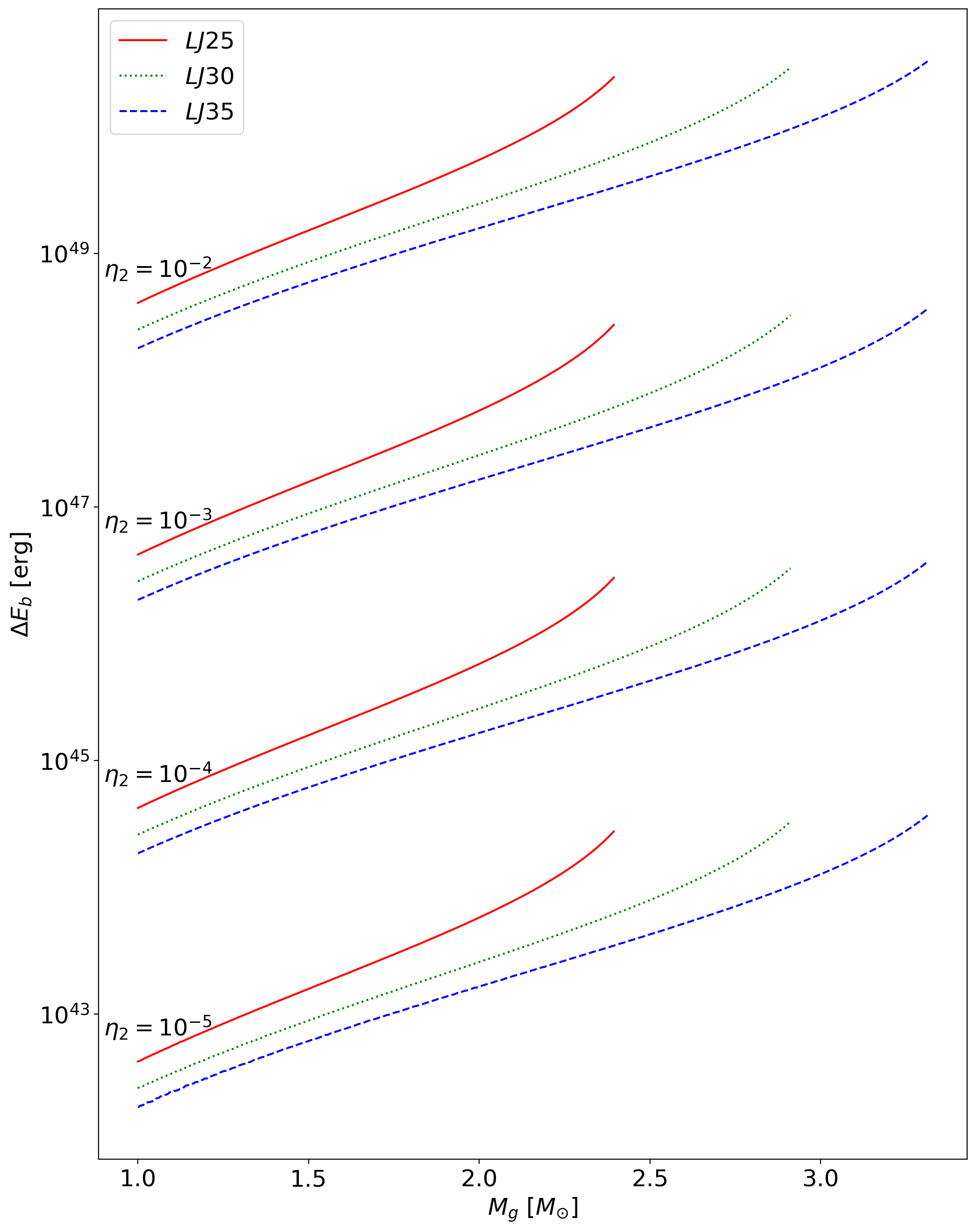}
    \caption{Same as Figure~\ref{fig:n=0_deltaEb}, except for the anisotropic model $\Pi=-\eta_2 rdp_{\rm r}/dr$.}
    \label{fig:n=1_deltaEb}
\end{figure}

\begin{figure}
	\includegraphics[width=0.78\columnwidth]{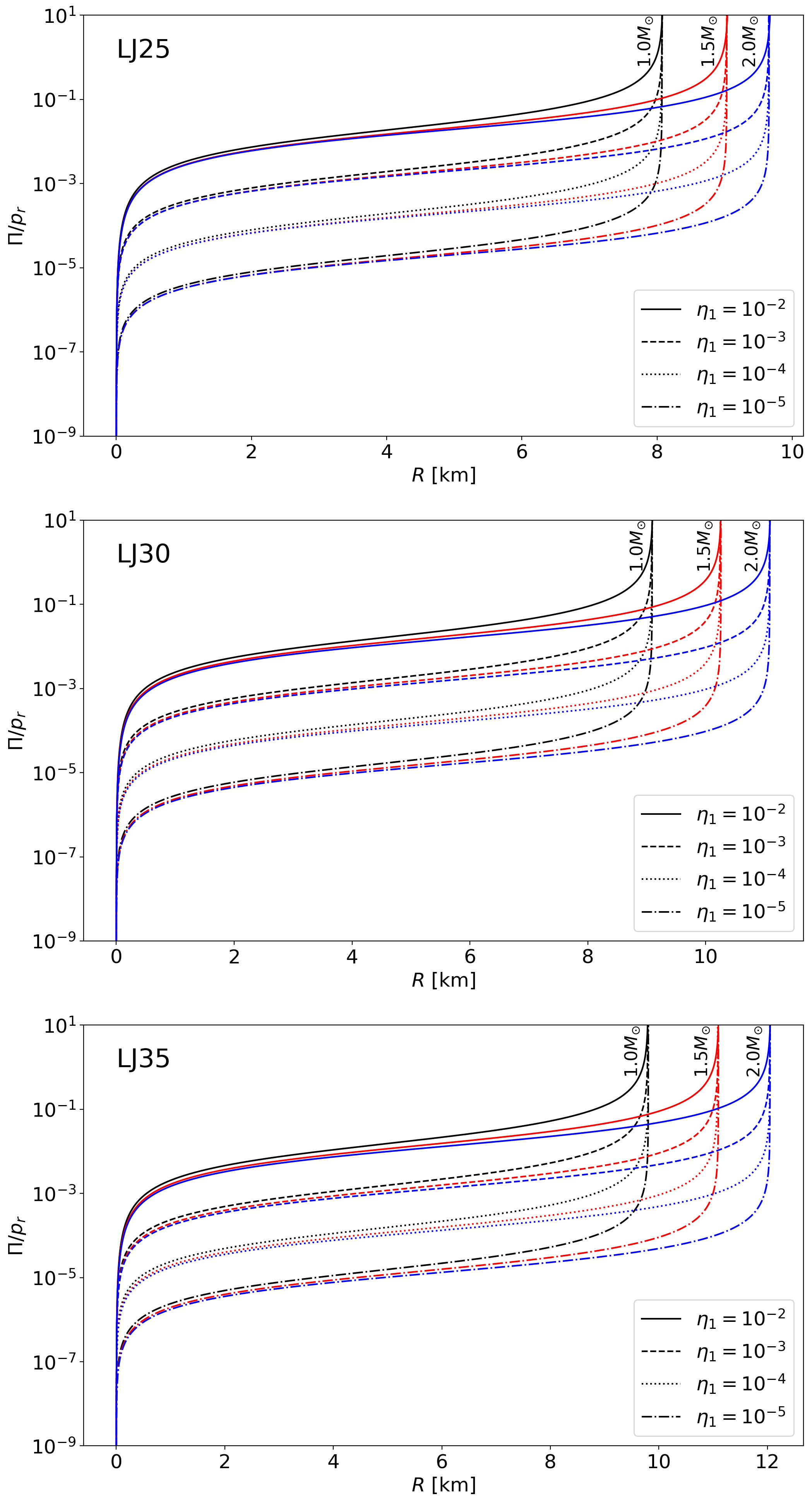}
    \caption{The value of $\Pi/p_{\rm r}$ as a function of radius, with the anisotropic model $\Pi=-\eta_1R_1dp_{\rm r}/dr$. Three sub-graphs correspond to three choices of EOS listed in table~\ref{tab:EOS}. Different line styles correspond to different values of $\eta_1$. The lines in the same line style with different colors correspond to different gravitational masses $M_{\rm g}$, which are labelled in the graph.}
    \label{fig:n=0_sigmatop}
\end{figure}

\begin{figure}
	\includegraphics[width=0.78\columnwidth]{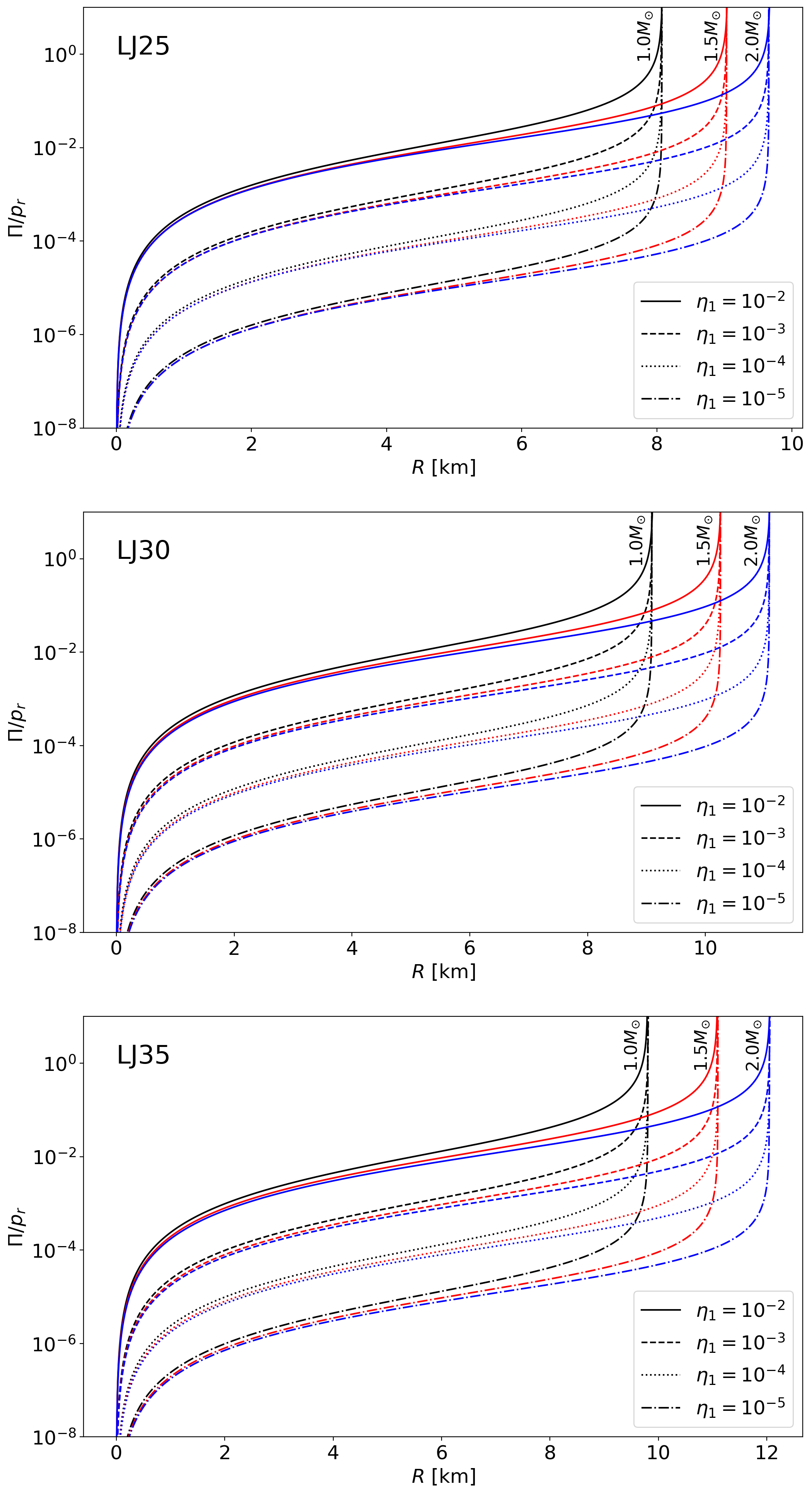}
    \caption{Same as Figure~\ref{fig:n=0_deltaEb}, except for the anisotropic model $\Pi=-\eta_2 rdp_{\rm r}/dr$.}
    \label{fig:n=1_sigmatop}
\end{figure}

The main results of our calculations are shown in Figure~\ref{fig:n=0_deltaEb}, \ref{fig:n=1_deltaEb},  \ref{fig:n=0_sigmatop} and \ref{fig:n=1_sigmatop}. Figure ~\ref{fig:n=0_deltaEb} and \ref{fig:n=1_deltaEb} show the difference of binding energy as a function of gravitational mass, implying the possible free energy the SSs may release via starquakes with different values of $M_{\rm g}$, $\eta_{1,2}$ and different equations of states. Figure \ref{fig:n=0_sigmatop} and \ref{fig:n=1_sigmatop} show the value of $\Pi/p_{\rm r}$ as a function of radius, which measures the local anisotropy within the stars with different values of $M_{\rm g}$, $\eta_{1,2}$ and different equations of states.

From Figure \ref{fig:n=0_deltaEb} and \ref{fig:n=1_deltaEb}, we can see that for larger mass and larger anisotropy (i.e. larger $\eta$), the potential free energy is larger. And given the same condition, LJ25 have the largest free energy, which means softer EOS tends to have larger potential free energy. And two anisotropic models have similar trends and shapes.

Figure \ref{fig:n=0_sigmatop} and \ref{fig:n=1_sigmatop} show that the higher the parameter $\eta_{1,2}$ is, the higher the $\Pi$ is, implying that the dimensionless constant $\eta_{1,2}$ does represent the local anisotropy. It can be seen that anisotropy or the difference of pressure is close to zero near the center of the star, and grows higher with larger radius. Three EOS models and two anisotropic models all have very similar trends and shapes.

From Figure~\ref{fig:n=0_deltaEb} and \ref{fig:n=1_deltaEb}, it is shown that for the model $\Pi=-\eta_1 R_1 \frac{dp_{\rm r}}{dr}$ with $\eta_1 = 10^{-4}-10^{-3}$ or for the model $\Pi=-\eta_2 r dp_{\rm r}/dr$ with $\eta_2=10^{-4}-10^{-3}$, the difference of binding energy ${\rm\Delta} E_{\rm b}$ is comparable to the typical energy of giant flare $\sim10^{44-47}$erg. From Figure~\ref{fig:n=0_sigmatop} and \ref{fig:n=1_sigmatop}, we can see that under these situations, the absolute value of ratio of $\Pi=p_{\rm t}-p_{\rm r}$ to $p_{\rm r}$ is approximately $10^{-5}-10^{-3}$.

\section{Discussions and conclusions}

The free energy of a SS would come from a release of the gravitational energy and the strain energy during starquakes, and extremely high magnetic fields might not be necessary in the case of SSs in order to understand various bursting events in astrophysics. The value of this free energy can be estimated as the difference of binding energy between the star with $\eta\neq0$ and $\eta=0$, where $\eta$ is a constant that measures the strength of local anisotropy, and $\eta=0$ means the star is isotropic and has no strain energy. In this paper, we calculate this kind of free energy of SSs in general relativity, and find that a small degree of anisotropy ($\Pi / p_{\rm r}\sim 10^{-4}$) can account for a large amount of free energy, comparable to the typical energy of giant flares ($\sim10^{44-47}$erg), as has already been illustrated in Newtonian gravity~\citep{2006MNRAS.373L..85X}.

Since we can not determine the anisotropic model on physical grounds from first principle, we choose two heuristic models by guess in this paper, $\Pi=-\eta_1 R_1 dp_{\rm r}/dr$ and $\Pi=-\eta_2 r dp_{\rm r}/dr$. Though we don't know the true form of the anisotropic model, these two toy models can at least show how anisotropy influence the free energy qualitatively. The influence of the anisotropy on the modified TOV equations is through $\Pi=p_{\rm t}-p_{\rm r}$, which only appears in equation \eqref{eq:tov3}. So, if the values of $\Pi$ within the star of two anisotropic models are similar, the value of free energy should be close too. Take the two models in the paper as an example. In the model $\Pi=-\eta_2 r dp_{\rm r}/dr$, there is a dimensionless constant $\eta_2$ which measure the intensity of anisotropy. In the model $\Pi=-\eta_1 R_1 dp_{\rm r}/dr$, we use the typical radius of pulsars $R_1=10\rm km$ to define a dimensionless constant $\eta_1$. Since we have two dimensionless constants $\eta_1$ and $\eta_2$ which measure the anisotropy, we can compare them. From figure \ref{fig:n=0_sigmatop} and \ref{fig:n=1_sigmatop}, we can see that, when $\eta_1$ and $\eta_2$ have the same order of magnitude, $\Pi/p_{\rm r}$ also have the same order of magnitude, so is the value of the free energy ${\rm\Delta} E_{\rm b}$. For $\eta_1\sim 10^{-3}$ and $\eta_2\sim 10^{-3}$, $\Pi/p_{\rm r}$ is around $10^{-4}-10^{-3}$ in the most part of the star except the center and the surface. Furthermore, as long as the anisotropic model can let $\Pi/p_{\rm r}$ to be over $10^{-4}-10^{-3}$ in the most part of the star, the free energy the star could release via starquakes can be over $10^{46}$erg, comparable to that of the giant flares. And from Figure~\ref{fig:n=0_deltaEb}, \ref{fig:n=1_deltaEb},  \ref{fig:n=0_sigmatop} and \ref{fig:n=1_sigmatop}, we can roughly guess that the increase of an order of magnitude in $\Pi/p_{\rm r}$ could make the free energy ${\rm\Delta} E_{\rm b}$ increase by two orders of magnitude.

However, it's still not clear that how the potential free energy can be transformed into radiation via starquake, so there aren't many things we can say about the details of the energy release process. The starquakes may created a self-induction electric field \citep{2015ApJ...799..152L}, which could initiate avalanches of pair creation in the magnetosphere and accelerate particles, inducing high-energy bursts \citep{2002ApJ...574..332T,2006MNRAS.373L..85X}. Since the radiation is starquake induced, it should have some characteristics of quakes. Following the law of seismology, if small quakes happen frequently, no big quakes would happen, but a giant quake may occur after long-time silence. And the glitches of pulsars may occur as X-ray transients, especially for the old SSs \citep{2006MNRAS.373L..85X}. And since it's very difficult to get the anisotropic model on physical ground from first principle, we could only guess some heuristic toy models, which are required to satisfy some conditions to make sure the results physically acceptable. All the factors that may have influence on the anisotropy, such as magnetic fields and relativistic nuclear interaction, are described roughly by the dimensionless constant $\eta$ which represents the magnitude of local pressure anisotropy. We also ignore the impact of rotation, since it only has minor influence on the gravitational mass, the change in $M_g$ due to rotation is less than one percent in normal cases \citep{2022MNRAS.509.2758G}, which means the rotation can be neglected when we focus on the free energy.

It's shown that the huge free energy ($\geq 10^{46}$ erg) could be released in SS via starquakes, even with very small anisotropy ($\eta \leq 10^{-4}$). This kind of free energy may be related to $\gamma$-ray bursts, fast radio bursts and soft $\gamma$-ray repeating sources,  without the need of extremely high magnetic field ($\sim 10^{15}$ G). There are many improvements can be done with this model in the future, since it's now only a phenomenological model with qualitative estimation of possible free energy in anisotropic SS. More detailed models on the starquakes and the process of energy transformation need to be built to give prediction on the observations, such as spectrum of radiation or signals of gravitational waves.

\section*{Acknowledgements}

The authors would like to thank those involved in the continuous discussions in the pulsar group at Peking University. This work is supported by the National SKA Program of China (2020SKA0120100).
E. Zhou is supported by NSFC Grant NO. 12203017.



\bibliographystyle{mnras}
\bibliography{RAA-2023-0178.R2} 


\bsp	
\label{lastpage}
\end{document}